\documentclass[aps,prl,reprint,superscriptaddress
]{revtex4-2}

\usepackage{hyperref}
\usepackage{graphicx}
\usepackage{amsfonts} 
\usepackage{thmbox} 
\usepackage{empheq} 
\usepackage{version}
\usepackage{times}
\usepackage{subfigure}
\usepackage{textcomp}
\usepackage{makeidx}
\usepackage{enumerate}
\usepackage{amsmath}
\usepackage{fancyhdr}
\usepackage[top=20mm, bottom=30mm, left=15mm, right=15mm]{geometry}
\usepackage{color}
\newcommand{\red}[1]{{\color{red}{#1}\normalcolor}}

\newcommand{\beq}{\begin{equation}}     \newcommand{\eeq}{\end{equation}}
\newcommand{\beqa}{\begin{eqnarray}}    \newcommand{\eeqa}{\end{eqnarray}}
\newcommand{\bde}{\begin{description}}  \newcommand{\ede}{\end{description}}
\newcommand{\ben}{\begin{enumerate}}    \newcommand{\een}{\end{enumerate}}
             
\newcommand{\noi}{\noindent\mbox{}}

\newcommand{\kT}{{k_{\rm B}T} } 

\newcommand{\eqn}[1]{\beq{ #1 }\eeq}

\newcommand{\inv}[1]{{\frac{1}{#1}}}

\newcommand{\inRbracket}[1]{{\left({#1}\right)}}
\newcommand{\inSbracket}[1]{{\left[{#1}\right]}}

\newtheorem[L]{thm}{Theorem}[section]
\newtheorem{cor}[thm]{Corollary}
\newtheorem{theorem}{{\sf Assertion :}}[section] 
\newtheorem{definition}{\sf Definition} 
 
\newtheorem{lemma}[theorem]{Lemma}
\newcommand{\blem}{\begin{lemma}}  
\newcommand{\elem}{\end{lemma}}  
\newcommand{\bpr}{\begin{proof}}  
\newcommand{\epr}{\end{proof}} 
\newcommand{\bdefine}{\begin{definition}}  
\newcommand{\edefine}{\end{definition}}  
\newcommand{\bcor}{\begin{cor}} 
\newcommand{\ecor}{\end{cor}}  
\newcommand{\bprop}{\begin{example}[Property]}  
\newcommand{\eprop}{\end{example}}  
\newcounter{formulaire}
\newcommand{\beqf}{\addtocounter{formulaire}{1}\begin{equation}}
\newcommand{\eeqf}{\tag{R \arabic{formulaire}}\end{equation}}
\newcommand{\beqaf}{\addtocounter{formulaire}{1}\begin{equation}\begin{array}{rcl}}
\newcommand{\eeqaf}{\end{array}\tag{R \arabic{formulaire}}\end{equation}}

\newcommand{\modif}[1]{{\null{#1}}}
\newcommand{\mydif}[1]{{\null{#1}}}
\newcommand{\repere}[1]{{\red{}}}

\begin{document}


\title{Memory Through a Hidden Martingale Process in Progressive Quenching }


\author{Charles Moslonka}
\affiliation{Laboratoire Gulliver - UMR 7083,  ESPCI Paris, France}
\affiliation{ENS-Paris-Saclay, Cachan, France}
\author{Ken Sekimoto}
\affiliation{Laboratoire Gulliver - UMR 7083,  ESPCI Paris, France}
\affiliation{Laboratoire Mati\`eres et Syst\`emes Complexes -UMR 7053, 
Universit\'e Paris Diderot, Paris, France}

\date{\today}

\begin{abstract}
Progressive quenching (PQ) is the stochastic process in which the system's degrees of freedom are sequentially fixed. While such process does not satisfy the local detailed balance, it has been found that the some physical observable of \mydif{a complete spin network} exhibits the martingale property. 
We studied system's response to the perturbation given at intermediate stages of the PQ.
The response at the final stage reveals the persistent memory, and we show that this persistence is a direct consequence of the martingale process behind. Not only the mean response, the shape of the probability distribution at the stage of perturbation is also memorized.  
Using the hidden martingale process we can predict the final bimodal distribution from the early-stage unimodal distribution \repere{2-6}\modif{in the regime where the unfrozen spins are paramagnetic}.
 We propose a viewpoint that the martingale property is a stochastic conservation law which is supported behind by some stochastic invariance. 
\end{abstract}


\maketitle

\section{Introduction}
The theory of linear response (Nakano-Kubo-Greenwood) has been established since long time to describe how the system in thermodynamic equilibrium reacts to the past perturbations given to it. 
The microscopic time-reversal invariance of equilibrium, i.e., the detailed balance (DB) symmetry played there a crucial role to bring out the fluctuation-dissipation (FD) relationship as well as Onsager's reciprocity law \cite{Onsager-1931,Onsager-Casimir-review}. 
Much less is known about the dynamic response of the systems which are far from equilibrium, especially when the elementary processes do not satisfy the local detailed balance (LDB).

Recently, the Malliavin weighting \cite{Berthier-prl2007,Warren-Allen-prl2012}, which is a special case of Malliavin derivative \cite{Malliavin76}, has been introduced to study the dynamic response of stochastic systems undergoing general Markovian process without assuming the LDB. 
In the present paper we study this type of general response especially when the system's dynamics exhibits the martingale property.
\repere{1-3}\modif{The martingale property means that an observable of the system undergoing stochastic process, \null{say $\hat{m}_T$} with $T$ being the time, evolves such that the conditional expectation of \null{$\hat{m}_{T+1}$ at time $T+1$} remains equal to \null{$\hat{m}_T$ under} the given history of the system up to $T$: 
\beq \label{eq:def-MG}
E[\hat{m}_{T+1} | \mathcal{F}_T]=\hat{m}_T, 
\eeq
where $E[X|\mathcal{F}_T]$ means to take the conditional expectation of $X$ given the history up to $T,$ and $\hat{m}_T$ is determined by $\mathcal{F}_T.$}

The background of this study is the following. 
We have studied what we call the progressive quenching (PQ) in which we fix progressively and  cumulatively, a part of system's degrees of freedom \cite{PQ-KS-BV-pre2018}. 
\repere{2-8}\modif{This procedure is reminiscent of the greedy algorithms.\footnote{This algorithm makes a sequence of choices which are in some way the best available and this never goes back on earlier decisions. See \cite{greedy-curtis} and the references cited therein.} }
More concretely, we focused on a totally connected Ising spins and fixed one spin after another while equilibrating the unfixed part of the spins every time we fix a single spin. 
If we regard the number of fixed spins $T$ as the discrete time, the distribution of the spin's fixed magnetization showed a sign of a long term memory. But at that time we had no good idea to quantify this memory as this quenching process breaks the LDB, and the FD relationship is not applicable.
  On the other hand, if we regard the equilibrium average of the unfixed spins after fixation of $T$-th spin (the equilibrium mean spin, for short, denoted by \null{$\hat{m}^{\rm (eq)}_{T}$}) as a stochastic process, it is found to have the martingale property up to small finite-size corrections, \mydif{which is essentially Eq.(\ref{eq:def-MG})} \cite{PQ-KS-BV-pre2018}. 

Having come to know the Malliavin weighting \cite{Berthier-prl2007,Warren-Allen-prl2012}, we retook {the PQ} problem and directly analyzed its response to the external field perturbations using the approach of Malliavin weighting. We found that the long memory of {the PQ} is a direct consequence of the martingale property it contains.
Below we focus on the response of the total magnetization in the final state when all the spins have been fixed. 

In the next section (\S\ref{sec:define}) 
 we first setup the model spin system and  define {the protocol} of progressive quenching under external perturbing field. 
Then we describe the response of the total magnetization in the final stage to the perturbing field  (\S\ref{sec:Results}). 
First we briefly recapitulate the previous result \cite{PQ-KS-BV-pre2018}) in \S\ref{subsec:unperturbed}. 
Then in \S\S\ref{subsec:Malliavin-weight}
 we take the approach of the Malliavin weight \cite{Berthier-prl2007,Warren-Allen-prl2012} adapted to the present PQ model. We calculate the response of the probability distribution of the total magnetization.
In \S\S\ref{subsec:OST} we focus on the response of a mean value of the total  magnetization, where the relevance to the martingale property is highlighted.
The power of the martingale property of {$\hat{m}^{\rm (eq)}_{T}$} is demonstrated when we use this to  predict the final distribution of total magnetization itself, not only its average (\S\S\ref{sec:Prediction}).
  In the concluding section \S\ref{sec:Conclusion} 
we formulate our core result in more general terms of discrete- and continuous-time stochastic processes. By this framework we will assert that, when a physical observable of a system possesses the martingale property,  this property acts as a kind of stochastic conservation law, 
  causing a long-term memory in the system's response, just like the true conservation laws played important roles in the response theory of the equilibrium systems through the emergence of   hydrodynamic modes, either diffusive or propagative \cite{MPP1972}. Also we will remark that, 
at least in the case of PQ, the stochastic conservation law is supported behind by a stochastic invariance property, \mydif{i.e., the invariance on average} .

\section{Setup of model and protocol\label{sec:define}}
\begin{figure}[h!!]
\centering 
\subfigure[\null ]
{   \label{fig:intro-a}  \includegraphics[width=3.3cm]{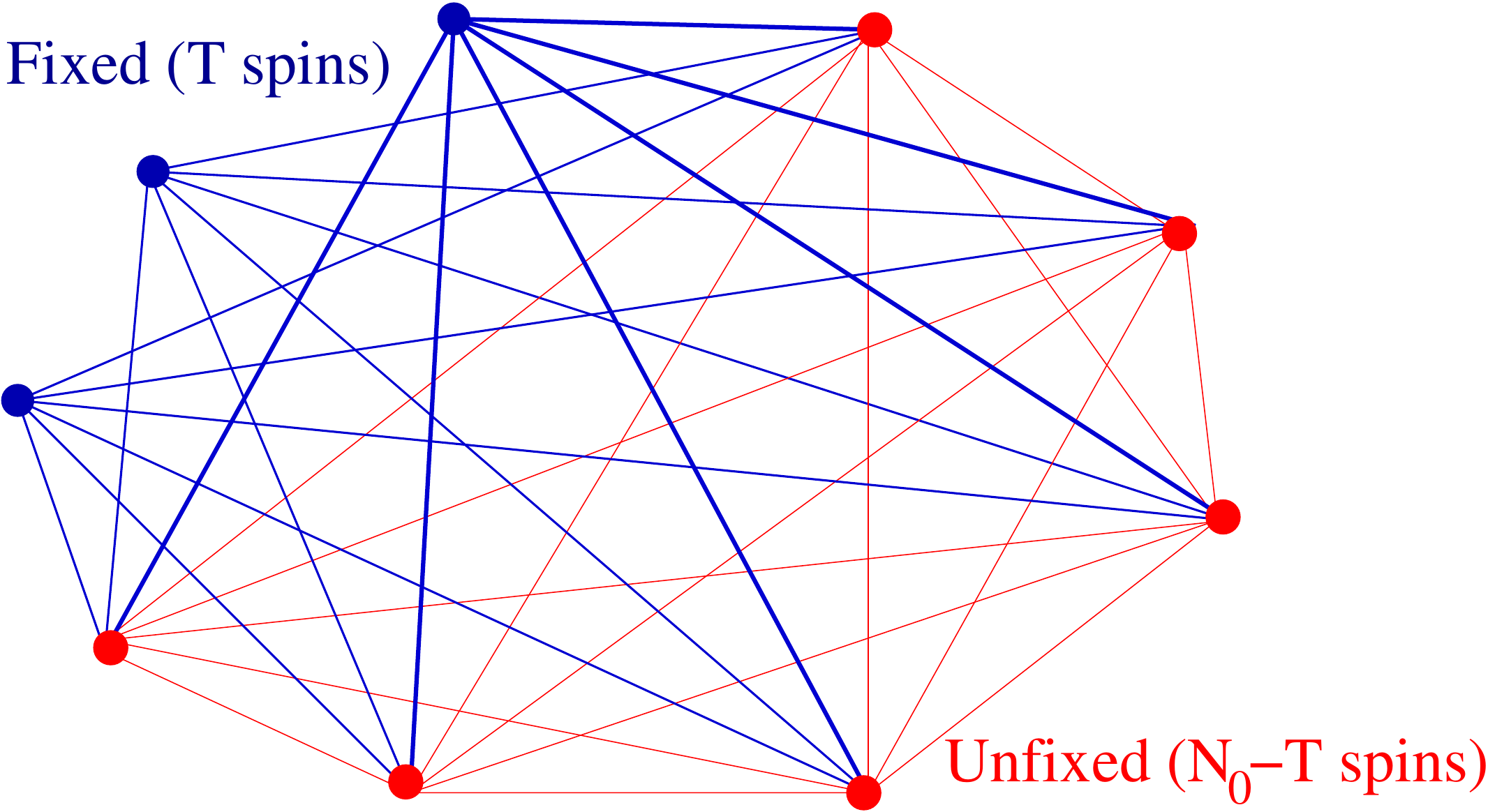}} 
\hspace{0.5cm}
\subfigure[\null ]
{   \label{fig:intro-b}  \includegraphics[width=3.3cm]{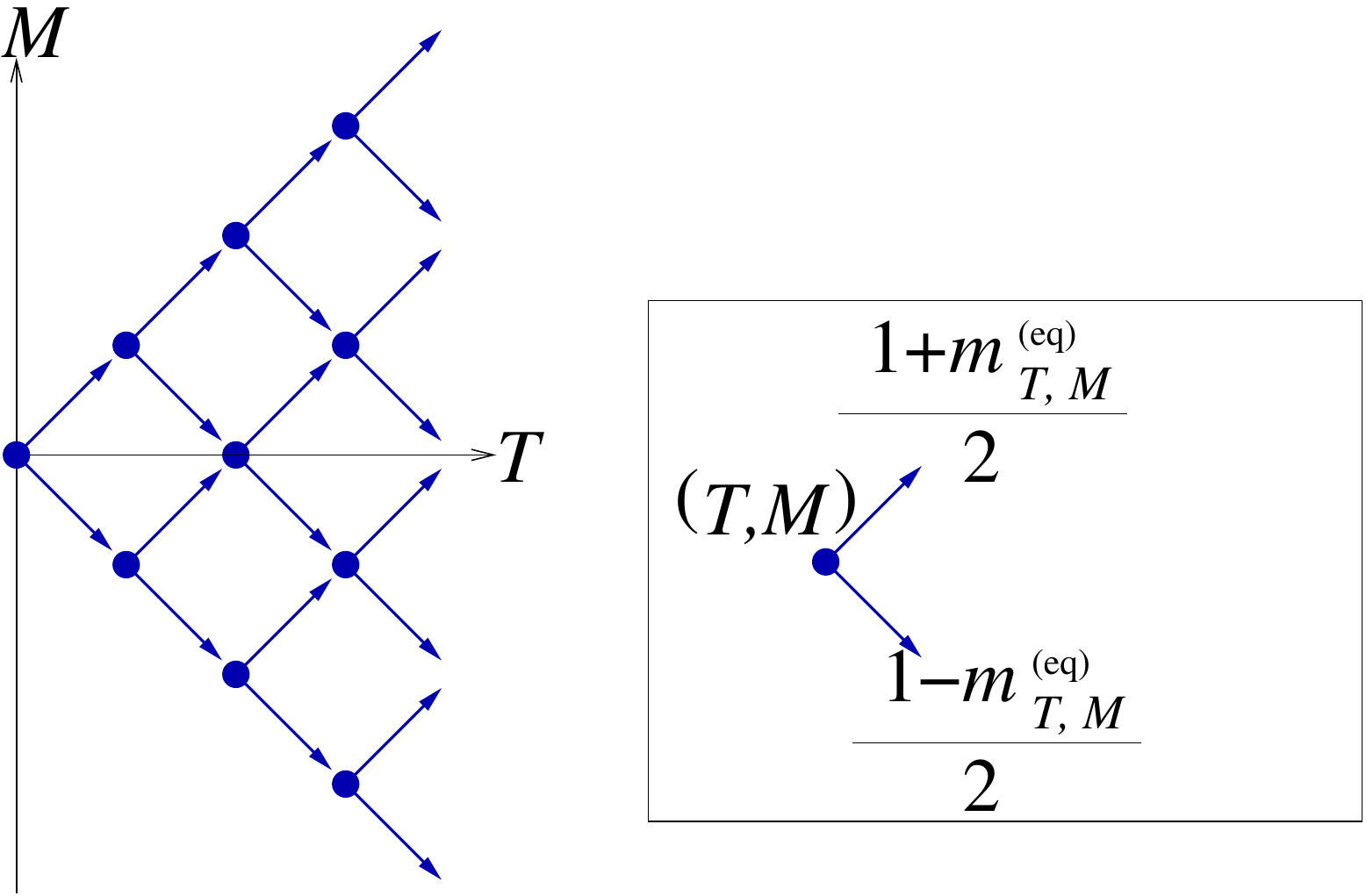}} 
\caption{(a) In the complete network of $N_0$ spins, $T$ spins have been fixed and there remain $N_0-T$ unfixed spins.
(b) PQ process of a complete spin network is a Markovian process on the 2D directed \mydif{network on the integer} lattice coordinated by $T$ and $M=\sum_{k=1}^T s_k.$  Those lattice points which are not visited are masked.} 
\label{fig:intro} 
\end{figure}

\paragraph{Globally coupled spin model : \label{subsec:GPS}}
We consider the ferromagnetic Ising model on a complete network, that is, the model in which any one of the spins interacts with all the other spins with equal coupling constant, {}{$j_0/N_0,$ where $N_0$ is the total number of spins.} 
We mean by the stage-$T,$ or simply $T,$ that there are $T$  spins that \mydif{have been} fixed, see Fig.\ref{fig:intro-a} for illustration.
When $N=N_0-T$ spins are unfixed  
under a field $h=h_T+{h_{\rm ext}}$, we use the energy function, 
\begin{equation}\label{eq:HTM}
\mathcal{H}_{T,M}=-\frac{j_0}{N_0} \sum_{T+1\leq i<j\leq N_0} s_i s_j \mydif{-}(h_T +{h_{\rm ext}})\sum_{i=T+1}^{N_0}s_i   ,
\end{equation}
where each spin $s_k$ takes the value $\pm 1.$
The field on the unfixed spins consists of two parts: One is $h_T:= \mydif{+}\frac{j_0}{N_0}M$ 
which is the ``\mydif{quenched} molecular field'' due to those fixed spins, $\{s_{1},\ldots,s_{T}\},$  where the total fixed magnetization is $M=\sum_{k=1}^{T}s_k$ 
and we have relabelled the spins for our convenience.
 The other part, ${h_{\rm ext}},$ is the genuine external field to perturb the process of PQ. 
 In the absence of perturbation we set ${h_{\rm ext}}=0.$
For the later use we introduce $m^{\rm (eq)}_{T,M}$ as the canonical average 
of the unfixed spins with the probability weight $e^{-\mathcal{H}_{T,M_{}}/\kT}.$ This is, therefore, the function of $T$ and $M=\sum_{k=1}^{T}s_k.$ 
In order to see clearly the effect of fluctuations, we choose the coupling constant $j_0$ so that the starting system $T=0$ is at the critical point of the finite system, $j_{0,\rm crit}$ (for the details, see \cite{PQ-KS-BV-pre2018}). 
Hereafter we let $\kT=1$ by properly choosing the unit of temperature.
 
\paragraph{Progressive quenching: }
The protocol of PQ is the cycle of re-equilibration of the unfixed spins and the fixation of a single spin at $\pm 1$ with the probabilities $ (1\pm m^{\rm (eq)}_{T,M})/2,$ respectively, see Fig.\ref{fig:intro-b}, where $m^{\rm (eq)}_{T,M}$ was defined above.
Once a spin is fixed, its value is retained until the end of the \mydif{whole} process.
Below we will use the notation $\hat{M}_T$ when we regard $\modif{M_T}=\sum_{k=1}^T s_k$ as stochastic process versus $T$ starting from $\hat{M}_0=0.$ The process $\hat{M}_T$ is Markovian. 
PQ can, therefore, be represented as a stochastic graph of $\hat{M}_T$ vs $T$ on the 2D discrete lattice, where the domain of $M$ is practically limited by $|M|\le T$ for each $T$ ($0\le T\le N_0$) \mydif{and $M\equiv T\,(\mbox{mod 2})$, see Fig.\ref{fig:intro-b}.}. 

\paragraph{Mapping to transfer matrix formulation : }
\mbox{}
\repere{2-2}
Instead of simulating the path ensemble, which would cost $\mathcal{O}(2^{N_0})$ trials, 
we can solve the master equation for the distribution of $\hat{M}_T,$ which costs no more than an algebraic power of $N_0$. By definition of PQ the partition between the system and the external system (i.e. fixed spins) is not static. We can, nevertheless, reformulate the evolution as that of a super-system
which is adapted to the transfer matrix method: 
The stochastic process of $\hat{M}_{T}$ vs $T$ with $0\le T\le N_0$ is represented as the transfer of $(2N_0+1)$-dimensional vector, $\vec{P}^{(T)}=\{ P^{(T)}_M\}_{M=-N_0}^{N_0}.$ 
The initial state $\vec{P}^{(0)}$ is $P^{(0)}_0=1$ for $M=0$ and $P^{(0)}_M=0,$ otherwise. 
The transition from the stage $T$ to the next one can be represented by 
a transfer matrix, ${\sf W}^{(T+1\leftarrow T)},$ such that 
$P^{(T+1)}_{M}=\sum_{M'=-T}^T ({\sf W}^{(T+1\leftarrow T)})_{M,M'} P^{(T)}_{M'}$  \mbox{or, in vector-matrix notation, } $\vec{P}^{(T+1)}={\sf W}^{(T+1\leftarrow T)} \vec{P}^{(T)}
$
for $0\le T\le N_0-1.$ 
The component of the matrix, $({\sf W}^{(T+1\leftarrow T)})_{M',M},$ is the conditional probability that the fixation of the $(T+1)$-th spin makes the total fixed magnetization change from $M$ to $M'.$ 
By definition of PQ the only non-zero components of ${\sf W}^{(T+1\leftarrow T)}$ are $({\sf W}^{(T+1\leftarrow T)})_{M\pm 1,M}$ with $|M|\le T$ and $M\equiv T\,(\mbox{mod }2).$
The transitions in the absence of perturbation (i.e. with $T\neq T_0$) gives $({\sf W}^{(T+1\leftarrow T)})_{M\pm 1,M}= (1\pm m^{\rm (eq)}_{T,M})/2$ 
corresponding to the fixation of the spin, $\hat{s}_{T+1}=\pm 1,$ respectively.
Using this notation, the final probability distribution of the total magnetization $M_{N_0}$ in the absence of the perturbation reads,
\beq \label{eq:PIeq} \vec{P}^{(N_0)}={\sf W}^{(N_0\leftarrow N_0-1)}\cdots {\sf W}^{(1\leftarrow 0)}\vec{P}^{(0)}.
\eeq
 Another key stochastic process is the mean equilibrium spin {$\hat{m}^{\rm (eq)}_{T}\equiv m^{\rm (eq)}_{T,\hat{M}_T}$}. \modif{
\repere{1-3} 
As was mentioned in the Introduction we have previously shown its martingale property (cf. Eq.(\ref{eq:def-MG})), and the consequence of Doob's optional sampling theorem (OST) \cite{OST-book}. 
\beq \label{eq:meq-MG}
E[\hat{m}^{\rm (eq)}_{T'} | \mathcal{F}_T]=
\hat{m}^{\rm (eq)}_T
 +\mathcal{O}((T'-T)/{N_0}^{2}), \quad T'>T,
\eeq
 where $\mathcal{F}_T\equiv \{\hat{M}_0,\hat{M}_1,\ldots, \hat{M}_{T}\}$ 
 and $E[\hat{M}_{T+1}-\hat{M}_T|\mathcal{F}_T]=\hat{m}^{\rm (eq)}_{{M}_T}$
  \cite{PQ-KS-BV-pre2018}.  Because $\{\hat{M}_T\}$ is a Markov process, we hereafter replace
this condition $\mathcal{F}_T$ by $\hat{M}_T.$ }
\modif{\repere{1-1} The martingale process $\{\hat{m}^{\rm (eq)}_T\}$ is {\it hidden} in the sense that the main observable, $\hat{M}_T,$ is not martingale by itself, see more discussion in \S\ref{sec:Conclusion}. }  

\paragraph{Application of the perturbation:}\mbox{}
{In the next section we will study the influences of} the external field perturbation ${h_{\rm ext}}$ which is applied uniquely at the stage-$(T_0-1).$ 
That is, in the presence of $h_{\rm ext}+h_{T_0-1},$ where  $h_{T_0-1}$ is the quenched molecular field by the fixed spins, we re-equilibrate $N_0-(T_0-1)$ spins before fixing the $T_0$-th spin. 
\repere{2-2} If the external field is applied at the stage-$(T_0-1),$ the matrix ${\sf W}^{(T_0\leftarrow T_0-1)}$ should be modified; we denote the corresponding transfer matrix by ${\sf W}_{h_{\rm ext}}^{(T_0\leftarrow T_0-1)}.$ The perturbed process and the resulting final distribution, $\vec{P}^{(N_0)}_{h_{\rm ext}}$ reads,
\beqa \label{eq:PIh}
\vec{P}^{(N_0)}_{h_{\rm ext}}
&=&{\sf W}^{(N_0\leftarrow N_0-1)}\cdots {\sf W}^{(T_0+1\leftarrow T_0)}
\cr &&
{\sf W}_{h_{\rm ext}}^{(T_0\leftarrow T_0-1)}
{\sf W}^{(T_0-1\leftarrow T_0-2)}\cdots
{\sf W}^{(1\leftarrow 0)}\vec{P}^{(0)}.
\eeqa 
  The martingale \null{property of $\hat{m}^{\rm (eq)}_{T}$} \cite{PQ-KS-BV-pre2018} is, therefore, interrupted upon the transition from the stage-$(T_0-1)$ to the stage-$T_0$. 
From the stage-$T_0$ the martingale property of \null{$\hat{m}^{\rm (eq)}_{T}$} with $T\ge T_0$ holds {\it de nouveau} with the total fixed spin $\hat{M}_{T_0}$ being the new initial condition. The question is how the perturbation given to $\hat{M}_{T_0}$ propagates up to the final value ${\hat{M}_{N_0}}$ and how the martingale property of \null{$\hat{m}^{\rm (eq)}_{T}$} manifests itself in this propagation.

\section{Results\label{sec:Results}}
\subsection{Unperturbed evolution --- R\'esum\'e}
\label{subsec:unperturbed}
\mbox{}\modif{\repere{1-4}\repere{2-2}
We recapitulate very briefly our previous study, where no external perturbations were applied \cite{PQ-KS-BV-pre2018}. We only show the evolution of the probability distribution of $M_T$ which is relevant to the following analysis.
 Fig.\ref{fig:result-distribution}(a) shows the snapshots of the distribution of $M_T$
 for the system of $N_0=2^8$ spins. These have been obtained essentially by 
 interrupting the calculation of Eq.(\ref{eq:PIeq}) at the midpoint; 
 $\vec{P}^{(T)}={\sf W}^{(T\leftarrow T-1)}\cdots {\sf W}^{(1\leftarrow 0)}\vec{P}^{(0)}.$ }
\modif{\repere{2-6} 
The coupling parameter $j_0$ is on the single phase side, i.e., $j_0 \le j_{0,\rm crit}.$ But if 
$j_0$ is not far below the critical one, the distribution develops bimodal shape, as seen in 
Fig.\ref{fig:result-distribution}(a). On the other hand if $0\le j_0<j_0^* (< j_{0,\rm crit})$ with some 
threshold coupling $j_0^*$, then the peak remains unimodal until the final stage. For example,
with $j_0=0$ the $\vec{P}^{(T)}$ is a symmetric binomial distribution. Whether or not $\vec{P}^{(T)}$ develops bimodal profile depends on the relative importance of the memory of the early stages,
such as the value of $\hat{s}_1=\pm 1.$ The memory of these stages is kept tenaciously in any case, but it can be blurred by the noises if the system's (paramagnetic) susceptibility in the early stages is not  large enough. This qualitative explanation will become clearer later in terms of the hidden martingale (\S\ref{sec:Prediction}).}

\modif{\repere{2-6} 
We recall that the appearance of bimodal profile of $P^{(N_0)}$ is {\it not} the result of the first order transition: The system of unfrozen spins is in the single para-magnetic phase because the 
effective coupling among them, $j_{\rm eff}=(1-\frac{T}{N_0})j_{0, \rm crit},$ is below critical
for all ${\sf W}^{(T+1\leftarrow T)}$ ($1\le T\le N_0$). Note that only above critical coupling do we have the first order transition.
As the quench proceeds this coupling is weaken progressively, i.e. the system becomes warmer and warmer above the critical temperature. Therefore, although the spin-spin coupling is global, there is no cooperativity, i.e., the thermal fluctuation of $m^{\rm (eq)}_{T,\hat{M}_T}$ is always unimodal for the individual system. It is the ensemble of systems that can develop the bimodal statistics like in Fig.\ref{fig:result-distribution}(a). 
In fact our previous numerical studies (\cite{PQ-KS-BV-pre2018}, Fig.3(c)) indicated that the threshold coupling parameter $j_0^*$ mentioned above behaves in such way that the gap $|j_0^* - j_{0,{\rm crit}}|$  disappears for $N_0\to\infty.$ The last tendency is opposite to the mean-field picture of the first order transition in which the bimodal nature should be most pronounced in the infinite-size limit.}

\subsection{Sensitivity of final-state distribution to perturbations
\label{subsec:Malliavin-weight}}
\repere{2-2}
The response to the perturbation given at the stage-$(T_0-1)$ can be studied in two complementary ways like the Fokker-Planck versus Langevin dynamics.
In the present subsection 
we follow how the perturbation given to $\vec{P}^{(T_0)}$ is transferred to that in the final distribution $\vec{P}^{(N_0)}$ through (\ref{eq:PIh}). 
This approach, of Fokker-Planck type, is in line with the Malliavin weighting \cite{Berthier-prl2007,Warren-Allen-prl2012} when the perturbation is infinitesimal (see below).  
In the next subsection \S\S\ref{subsec:OST} we rather focus on the evolution of 
$\hat{M}_T$ from $T=T_0$ up to $T=N_0,$ similar to the Langevin equation but through the filter of the conditional expectation, $E[\hat{M}_T| M_{T_0}].$

The direct consequence of the perturbation at the stage-$(T_0-1)$ is the shift of 
the transfer matrix, $\Delta {\sf W}^{(T_0\leftarrow T_0-1)} \equiv {\sf W}_{h_{\rm ext}}^{(T_0\leftarrow T_0-1)}-{\sf W}^{(T_0\leftarrow T_0-1)}.$ As the result of propagation of the shift the final shift of the probability density reads,
\beqa \label{eq:DeltaP}
\vec{P}^{(N_0)}_{h_{\rm ext}}-\vec{P}^{(N_0)}
 &&=
{\sf W}^{(N_0\leftarrow N_0-1)}\cdots {\sf W}^{(T_0+1\leftarrow T_0)}
\cr && \!\!\!\!\!\!\!\!\!\!\!\!\!\!\!
\Delta {\sf W}^{(T_0\leftarrow T_0-1)}
 {\sf W}^{(T_0-1\leftarrow T_0-2)}\cdots{\sf W}^{(1\leftarrow 0)}\vec{P}^{(0)}.
\eeqa

In the case of the infinitesimal perturbing field, we deal with the linear response to $h_{\rm ext}$ and calculate, instead of (\ref{eq:DeltaP}), the {sensitivity} 
\beqa \label{eq:dPdh}
\left.\frac{\partial \vec{P}^{(N_0)}_{h_{\rm ext}}}{\partial h_{\rm ext}}\right.
&=&{\sf W}^{(N_0\leftarrow N_0-1)}\cdots {\sf W}^{(T_0+1\leftarrow T_0)}
\cr && \!\!\!\!\!
{\tiny \left.\frac{\partial {\sf W}_{h_{\rm ext}}^{(T_0\leftarrow T_0-1)}}{\partial h_{\rm ext}}\right.} 
 {\sf W}^{(T_0-1\leftarrow T_0-2)}\cdots{\sf W}^{(1\leftarrow 0)}\vec{P}^{(0)},
\eeqa
where the partial derivative with respect to $h_{\rm ext}$ should be evaluated at $h_{\rm ext}=0$ and 
 the only non-zero components of 
 ${\partial {\sf W}_{h_{\rm ext}}^{(T_0\leftarrow T_0-1)}}/{\partial h_{\rm ext}}$
  are  ${\partial ({\sf W}_{h_{\rm ext}}^{(T_0\leftarrow T_0-1)})_{M\pm 1,M}}/{\partial h_{\rm ext}}  =\pm \chi^{\rm (eq)}_{T_0-1,M}/2$ for $|M|\le T_0-1$
  with $\chi^{\rm (eq)}_{T,M}\equiv \partial m^{\rm (eq)}_{T,M}/\partial h_{\rm ext}$ being the susceptibility at the stage-$T$ under a molecular field, $h_T= \frac{j_0}{N_0}M.$  
The approach of Malliavin weighting \cite{Berthier-prl2007,Warren-Allen-prl2012} is essentially the path-wise expression of (\ref{eq:dPdh}), see Appendix \ref{app:B} for more detailed account. 
In Fig.\ref{fig:result-distribution} (b) we plotted the result in (\ref{eq:dPdh}) vs $M_{N_0}/N_0$  
of the system with the size $N_0=2^8=256.$ 
Depending on the stage of perturbation ($T_0=2^4=16$ or $T_0=2^7=128$)
the sensitivity qualitatively changes, see below.
\mbox{}
\begin{figure}[b] 
\includegraphics[width=16.2cm]{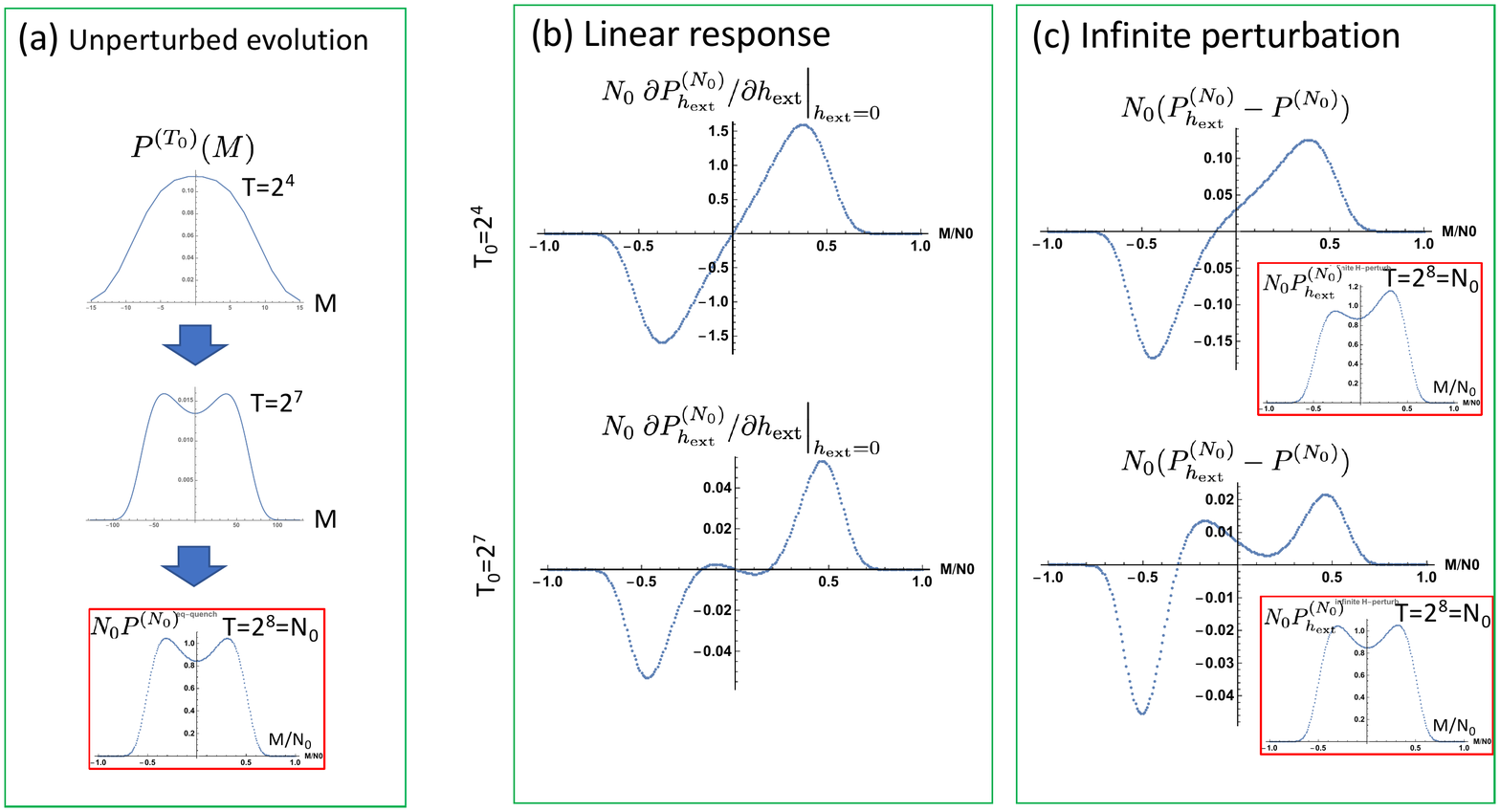}
\caption{
(a) The unperturbed evolution of the probability of the total fixed magnetization, $M$, at three different numbers of fixed spins, $T=2^4, 2^7$ and 
$T=2^8=N_0.$ (b) The linear response of the final distribution $P^{(N_0)}$ to the infinitesimal perturbations given at the different stages, $T_0=2^4$ (top) and $2^7$ (bottom). The horizontal axis is scaled by the system size.
(c) The response of the final distribution ${P}^{(N_0)}_{h_{\rm ext}}$ to the infinite perturbation given at the different stages, $T_0=2^4$ (top) and $2^7$ (bottom). The insets show the final perturbed distributions. 
}
\label{fig:result-distribution} 
\end{figure}

In the case of the infinite perturbing field $h_{\rm ext}=+\infty,$ we calculate
directly (\ref{eq:DeltaP}), where 
the transition rates upon the perturbed stage read $({\sf W}^{(T_0\leftarrow T_0-1)}_{h_{\rm ext}}) _{M+ 1,M}=1$ with $|M|\le T_0-1$ and all the remaining components of ${\sf W}^{(T_0\leftarrow T_0-1)}_{h_{\rm ext}}$ are zero. Therefore the only non-zero components of $\Delta {\sf W}^{(T_0\leftarrow T_0-1)}$ are $\Delta ({\sf W}^{(T_0\leftarrow T_0-1)})_{M\pm 1,M}=\pm(1-m^{\rm (eq)}_{T_0-1,M})/2$ for $|M|\le T_0-1.$

In Fig.\ref{fig:result-distribution}(c) we monitored $\vec{P}^{(N_0)}_{h_{\rm ext}}-\vec{P}^{(N_0)}$ vs $M_{N_0}$ as the response to the infinite perturbation, $h_{\rm ext}=+\infty.$ This response is qualitatively similar to the linear response of the distribution (Fig.\ref{fig:result-distribution}(b)),  except for a positive bias around $M_{N_0}= 0$ in the former case. We notice the two common trend for the both types of perturbation:  (i) The response is stronger when the perturbation is given at the early stage, which is 
contrasting to the equilibrium system for which the impact of perturbation should be strongest if it is given most recently, i.e. with the largest $T_0$. 
(ii) The profiles of the response reflects the distribution at the stage when the perturbations have been applied: If a perturbation is given when the unperturbed distribution of $M$ is still unimodal (i.g.  $T_0=2^4$), the density response in the final magnetization resembles to the $M$-derivative of the unimodal distribution at the stage-$T_0$. (Notice, however, that the width of distribution is ``magnified'' from $|M|\le T_0(=16)$ to the final one ranging over $|M|\lesssim 0.7\times N_0(\simeq 180)$.) 
Similarly, if the perturbation is given in the late stage (i.g. $T_0=2^7$), 
the final response resembles to the $M$-derivative of the bimodal distribution at $T_0.$
\modif{\repere{2-6}This trend (ii) suggests the presence of an underlying mechanism by which the individual realization of PQ keeps the memory of the stage when the perturbation is given. As noted in \S\ref{subsec:unperturbed} the possibility of first order transition is excluded. We will see later in \S\ref{subsec:OST} (especially Eq.(\ref{eq:EMN0})) that the origin of the memory is the (hidden) martingale property of $m^{\rm (eq)}_{T,\hat{M}_T}.$} 
\clearpage %
\subsection{Mean response of the final magnetization, $E[{\hat{M}_{N_0}}]$
\label{subsec:OST}}
We study the mean response of the total spin at the final stage, $E[{\hat{M}_{N_0}}],$
when an infinite perturbing field ($h_{\rm ext}=+\infty$) is applied at the stage-$(T_0-1),$ just before fixing the $T_0$-th spin. 
While this mean value $E[{\hat{M}_{N_0}}]$ can be calculated through (\ref{eq:PIh}),
here we will take a different approach; 
\beq \label{eq:EMN0-exact}
E[{\hat{M}_{N_0}}]=\sum_{M=-T_0}^{T_0} 
 E[{\hat{M}_{N_0}}|M_{T_0}=M]
 P^{(T_0)}_{h_{\rm ext}}(M),
\eeq
where $E[{\hat{M}_{N_0}}|M_{T_0}=M]$ is the conditional expectation.
By $({\sf W}^{(T_0\leftarrow T_0-1)}_{h_{\rm ext}}) _{M+ 1,M},$ which is
described in the last paragraph of \S\ref{subsec:Malliavin-weight}, 
$P^{(T_0)}_{h_{\rm ext}}(M)$ is the shifted copy of the previous stage, that is, $P^{(T_0)}(M+1)=P^{(T_0-1)}(M)$ for $|M|\le T_0-1$ and $P^{(T_0)}(-T_0)=0.$ Therefore,
for  $T_0$ not very large ($\ll N_0$) the calculation of $P^{(T_0)}_{h_{\rm ext}}(M)$ is a relatively light calculation. 
As for the conditional expectation $E[{\hat{M}_{N_0}}|M_{T_0}=M],$ if we use the martingale property of $m^{\rm (eq)}_{T,\hat{M}_{T}}$ for the unperturbed process $T_0\le T\le N_0$, we can show the compact result:
\beq\label{eq:EMN0}
E[{\hat{M}_{N_0}}|M_{T_0}=M]
=M+(N_0-T_0)\null{m^{\rm (eq)}_{T_0,M}} + \mathcal{O}({1}).
\eeq
Therefore, (\ref{eq:EMN0-exact}) reads finally
\beq\label{eq:EMN0-fin}
E[{\hat{M}_{N_0}}]=E[{\hat{M}_{T_0}}]+(N_0-T_0) 
E[m^{\rm (eq)}_{T_0,\hat{M}_{T_0}}] + \mathcal{O}({1}).
\eeq
Because the left hand side of (\ref{eq:EMN0}) is $\mathcal{O}(N_0)$, the error term of $\mathcal{O}(1)$ is negligible for $N_0\gg 1.$
Note that (\ref{eq:EMN0-fin}) does not require the calculation of transfer matrices beyond the stage-$T_0$. 
\begin{figure}[h]
\centering
{\includegraphics[width=6.2cm]{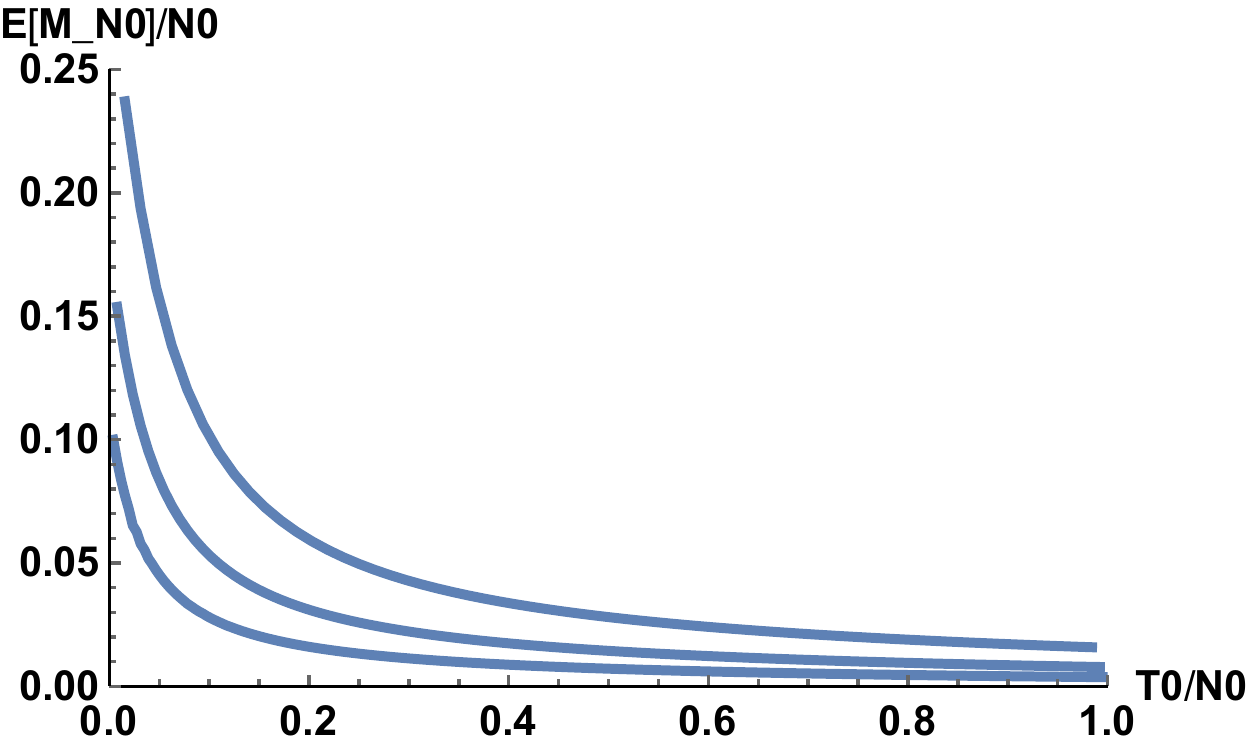}}
\caption{Mean response of the magnetization, $E[{\hat{M}_{N_0}}],$ 
to the perturbation $h_{\rm ext}=+\infty$ applied at the stage-$T_0.$
The system sizes. \null{The system size for each curve is  $N_0=2^6,2^7$ and $2^8,$ respectively, from top to bottom. }
 }
\label{fig:result-mean} 
\end{figure}

The relation (\ref{eq:EMN0}) 
comes out from a more general statement about the mean increment rate of $\hat{M}_T$:
$E[\frac{\hat{M}_{T}-M_{T_0}}{T-T_0}|M_{T_0}=M]= \null{m^{\rm (eq)}_{T_0,M}}
+\mathcal{O}((T-T_0)/{N_0}^2)$
for $T_0<T\le N_0.$ 
The derivation is given in Appendix \ref{sec:appendixA}, where we use 
the martingale property of $m^{\rm (eq)}_{T,\hat{M}_T}$ (see \ref{eq:meq-MG}).
The relation (\ref{eq:EMN0}) tells us that 
 the impact of perturbation is directly transmitted by the martingale observable, $m^{\rm (eq)}_{T,\hat{M}_T}.$ This opens the possibility to predict approximately the final distribution $P^{(N_0)}(M_{N_0})$ from the data at the stage-$T_0$ when the perturbation is given (see \S\ref{sec:Prediction} below) and then to understand better the result of \S\ref{subsec:Malliavin-weight}.
\modif{\repere{2-4} Because it is only in the expectation the mean increment rate, $\frac{\hat{M}_{T}-M_{T_0}}{T-T_0}|_{M_{T_0}=M},$ is kept constant over $T_0<T\le N_0,$  we call it the {\it stochastic conservation}.}

In Fig.~\ref{fig:result-mean} we plot the mean values of the final magnetization, $E[\hat{M}_{N_0}],$  
 The different curves in Fig.~\ref{fig:result-mean} correspond to the different system sizes, $N_0=2^6, 2^7$ and $2^8$. The both axes are rescaled by the system sizes. 
The formula Eq.(\ref{eq:EMN0-fin}) reproduces $E[\hat{M}_{N_0}]$ so well that the deviation from the full numerical results using  $P^{(N_0)}_{h_{\rm ext}}(M)$ is within the thickness of the curves.
That the mean response of the frozen spin, $E[\hat{M}_{N_0}]/N_0$ decreases with the system size $N_0$ is consistent with our previous observation in \S\ref{subsec:Malliavin-weight}, especially Fig.\ref{fig:result-distribution}(c). 

\subsection{Hidden martingale property predicts final distribution}\label{sec:Prediction}
\begin{figure}[h!!]
\centering
{\includegraphics[width=8.5cm]{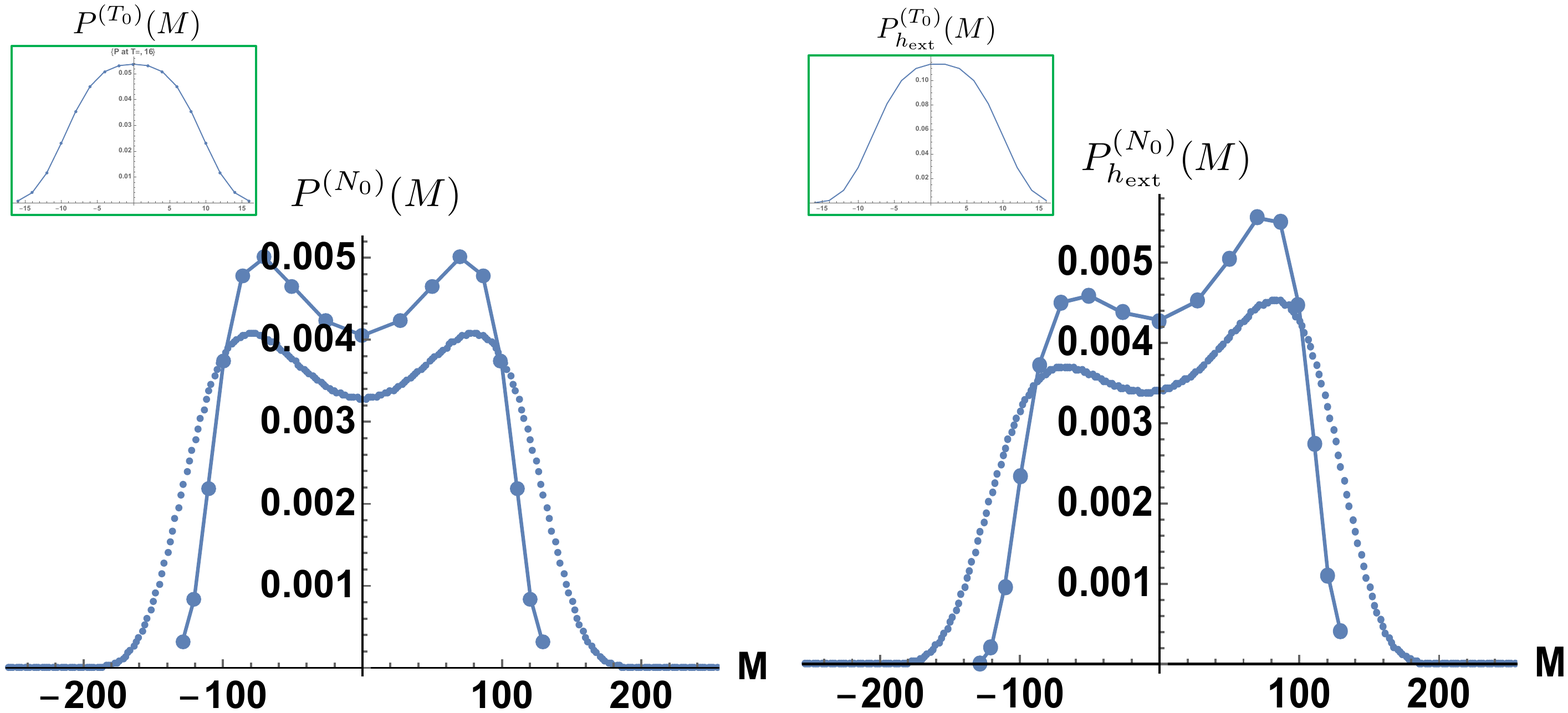}}
\caption{Comparison between the final distributions of $M_{N_0}$ predicted by the hidden martingale property (joined $T_0+1$ dots) with those by full numerical solution (dense dots)
for $h_{\rm ext}=0$ (left) and for $h_{\rm ext}=\infty$ (left) with $T_0=2^4$ and $N_0=2^8.$
In the figures the probability densities are rescaled so that their integral over $M$ be normalized to unity.  The figures in inset show the probabilities $P^{(T_0)}_{}$ (left) and
$P^{(T_0)}_{h_{\rm ext}}$ (right), respectively. Both are singly peaked but the latter is 
almost translocated by $\Delta M=+1.$}
\label{fig:fig3} 
\end{figure}

The \mydif{fluctuation} property of $m^{\rm (eq)}_{T,\hat{M}_{T}}$ adds something on top of 
(\ref{eq:EMN0}) when the system is large enough in the sense of $N_0\gg T_0.$
Starting from the condition $\hat{M}_{T_0}=M,$ the final magnetization
$\hat{M}_{T_0}$ should scatter around $E[{\hat{M}_{N_0}}|M_{T_0}=M],$ but its 
standard deviation should to be $\mathcal{O}((N_0)^{\inv{2}}),$ therefore, less dominant than the mean part, $(N_0-T_0) m^{\rm (eq)}_{T_0,M}=\mathcal{O}(N_0).$ 
This estimation of the standard deviation, $\mathcal{O}((N_0)^{\inv{2}}),$ is related to the so-called martingale central-limit theorem (see, for example, \S 3.3 of \cite{HandH}) together with the fact that 
$m^{\rm (eq)}_{T,\hat{M}_T}$ is non-extensive quantity of $\mathcal{O}(1).$ 
%
%
With the tolerance of $\mathcal{O}({N_0}^{\inv{2}})$ errors,
 Eq.(\ref{eq:EMN0}) leads, therefore, to a sort of geometrical optics approximation (\cite{feynman-optics} \S 27):
\beq\label{eq:geomopt}
\left.\hat{M}_{N_0}\right|_{M_{T_0}=M} =M+(N_0-T_0) \,m^{\rm (eq)}_{T_0,M}
+\mathcal{O}({N_0}^{\inv{2}}).
\eeq
This estimation in turn allows us to reconstruct the final probability distribution $P^{(N_0)}_{h_{\rm ext}}(M)$ versus $M,$ see Appendix \ref{sec:appendixC} for the detailed protocol.
In Fig.\ref{fig:fig3} we compare the final distributions of $\hat{M}_{N_0}$, one by the geometrical optics approximation and the other by the full numerical calculation of transfer matrix products. 
Naturally, the former method gives narrower distribution because this approximation ignores the broadening by the standard deviation, $\sim (256)^{\inv{2}}\simeq 16.$ 
Amazingly the geometrical optic approximation can nevertheless predict the positions of bimodal peaks very well from the data of unimodal distribution at the stage-$T_0.$
When $N_0$ and $T_0$ constitute the double hierarchy $1\ll T_0\ll N_0,$ our methodology will serve as a fine tool of numerical asymptotic analysis.
\modif{\repere{2-6} We have chosen the coupling $j_0$ at the critical one, $j_{0,\rm crit}$ because the predictability of bimodal distribution from unimodal stage looks impressive. Nevertheless, the tenacious memory given Eq.(\ref{eq:geomopt}) and the predictability as its consequence hold also for the weaker coupling with which the final distribution is unimodal. }

\section{Conclusion --- General argument\label{sec:Conclusion}}
\noi
\repere{1-1}
\repere{2-4} 
\modif{
We first summarize, using a general terminology, the mechanism by which the hidden martingale property gives rise to a tenacious memory of the process. We will use the notation which 
corresponds to the previous sections, such as $\hat{M}_T$ or $\hat{m}_T$, but we don't 
rely on the PQ model. 

Suppose that $\{\hat{M}_T\}$ ($0\le T\le N_0$) is a stochastic process with the discrete time $T$ and has the increment, $\hat{s}_{T+1}\equiv \hat{M}_{T+1}-\hat{M}_{T}.$ We assume that the probabilistic characteristics of $\hat{s}_{T+1}$ is determined by the history of $\{\hat{M}_t\}$ up to $t=T,$ and that its conditional expectation $\hat{m}_{T}\equiv E[\hat{s}_{T+1}|\mathcal{F}_T]$ is completely determined by the history up to $T,$ denoted by $\mathcal{F}_T.$ 
With only these settings we can verify that $\hat{R}_T\equiv \sum_{t=0}^{T-1} (\hat{s}_{t+1}-\hat{m}_t)$ is martingale, i.e., $E[\hat{R}_{T+1}|\mathcal{F}_T]=R_{T},$ the fact which is known as Doob-L\'evy decomposition theorem \cite{Williams1991,Doob1971}.
\repere{1-1} The martingale of our concern, however, is not this fact but we add another layer; we suppose that $\{\hat{m}_T\}$ is again martingale, that is, $E[\hat{m}_{T+1}|\mathcal{F}_T]=\hat{m}_{T}.$ This is why we call the latter the hidden martingale. 
 The outcome is that we have 
 \beq \label{eq:general-discrete}
 E[\hat{M}_{T}|\mathcal{F}_{T_0} ]=\hat{M}_{T_0}+ (T-{T_0}) \hat{m}_{T_0},\quad  T>T_0, 
 \eeq
which we can verify by following exactly the same argument as in Appendix\ref{sec:appendixA}
except that $\mathcal{O}((T-T_0)/{N_0}^2)$ in (\ref{eq:A1}) is omitted.

\repere{1-1} Eq.(\ref{eq:general-discrete}) tells how the hidden martingale property of $\{\hat{m}_T\}$ transmits  the memory of the past data without exponential or power-low decays. This relation is the general outcome of hidden martingale and has nothing to do with the origin of the hidden martingale. 
Especially, in our PQ model  the relation Eq.(\ref{eq:EMN0}) represents the tenacious memory  whether the distribution $P^{(N_0)}(M)$ is unimodal or bimodal.

For completeness, we also write down the continuous-time counterpart: 
Suppose that $\{\hat{M}_t\}$ ($0\le t\le t_0$) is a stochastic process with the continuous time $t$ and we denote the increment by $d\hat{M}_{t}\equiv \hat{M}_{t+dt}-\hat{M}_{t}.$ We assume that the probabilistic features of $d\hat{M}_{t}$ is determined by the history of $\{\hat{M}_\tau\}$ up to $\tau=t$ and its conditional expectation $\hat{m}_t\, dt \equiv E[d\hat{M}_{t}|\mathcal{F}_t]$ is completely determined by the history up to $t,$ denoted by $\mathcal{F}_t.$ 
Then by Doob-L\'evy decomposition theorem \cite{Williams1991,Doob1971} and the martingale central-limit theorem (see, for example, \S 3.3 of \cite{HandH}) allows to represent 
the stochastic evolution of $\hat{M}_t$ in the form of stochastic differential equation
\beq \label{eq:hMG-SDE}
d\hat{M}_t = \hat{m}_t \, dt + \hat{b}_t \cdot d\hat{W}_t,
\eeq
where the second term on the r.h.s. is an It\^o integral with a Wiener process, $\hat{W}_t.$
Now if we further suppose that $\{\hat{m}_t\}$ is martingale, then we have 
\eqn{ E[\hat{M}_{t}|\mathcal{F}_{t_0}]
=M_{t_0}+ (t-{t_0}) \hat{m}_{t_0}, \quad t>{t_0}
}
because $E[d\hat{M}_{t}|\mathcal{F}_{t_0}]=E[\hat{m}_{t}|\mathcal{F}_{t_0}] dt =\hat{m}_{t_0}dt$
holds for $t>t_0.$}

\modif{\repere{2-4} In \S\ref{subsec:OST} we called the formula of the type Eq.(\ref{eq:general-discrete}) the stochastic conservation law. This property leads to the lasting memory in the system's response. }
\null{In analogy with the (deterministic) physical conservation laws,} a far-fetched question would be if there is a kind of {stochastic invariance} behind the stochastic conservation, \null{just as many (deterministic) physical conservation laws are based on some invariance principle.}
\repere{1-2}\modif{In our setup the spin system the ``total molecular field'' on each unfrozen spin, which is the sum of the quenched molecular field $h_T$ and the interaction field from the other unfrozen spins, remains invariant upon the fixation of a spin (see Appendix C of  \cite{PQ-KS-BV-pre2018} for a mean-field argument).}

\begin{acknowledgments}
CM thanks the laboratory Gulliver 
at ESPCI for the encouraging environment to start the research. KS 
thanks Izaak Neri for fruitful discussions. KS benefits from the project JT of RIKEN-ESPCI-Paris 7.
\end{acknowledgments}

\appendix
\section{Derivation of Eq.(\ref{eq:EMN0})\label{sec:appendixA}}
The total fixed spins $\hat{M}_{T}$ at the stage-$T$ with $T_0<T\le N_0$ reads $\hat{M}_{T}=\hat{M}_{T_0}+\sum_{j=T_0+1}^T \hat{s}_{j},$
where $\hat{s}_{j}$ is the value of the spin which is fixed in the $j$-th quenching.
Taking the expectation of the above formula, i.e., $E[\hat{M}_{T}|\hat{M}_{T_0}=M_{T_0}]={M}_{T_0}+\sum_{j=T_0+1}^T E[\hat{s}_{j}|\hat{M}_{T_0}=M_{T_0}],$
we will focus on $E[\hat{s}_{j}|\hat{M}_{T_0}=M_{T_0}].$ For $T_0<T\le N_0$ the last quantity can be transformed as 
\beqa \label{eq:A1}
E[\hat{s}_{T}|\hat{M}_{T_0}=M_{T_0}]
&=&E\inSbracket{ E[\hat{s}_{T}|\hat{M}_{T-1}]\,|\hat{M}_{T_0}=M_{T_0}]}
\cr &=&
E\inSbracket{ m^{(\rm eq)}_{\hat{M}_{T-1}}|\hat{M}_{T_0}=M_{T_0}}
\cr &=&
m^{(\rm eq)}_{\hat{M}_{T_0}}+\mathcal{O}\inRbracket{\frac{T-T_0}{{N_0}^2}},
\eeqa
where, to go to the last line, we have used (\ref{eq:meq-MG}) with 
$(T',T)$ there being replaced by $(T,T_0)$ here, respectively. 
By choosing $T=N_0$ we arrive at Eq.(\ref{eq:EMN0}).

\section{Simple summary of Malliavin weighting\label{app:B}}
We explain the Malliavin weighting of \cite{Berthier-prl2007,Warren-Allen-prl2012}. 
The evolution of the probability distribution from the initial one to the finale one is given as the matrix-vector product like (\ref{eq:PIeq}) or (\ref{eq:PIh}) in the main text. 
These product can be regarded as the discrete path integrals because the different paths to reach the final state $\hat{M}_{N_0}$ form the initial one $\hat{M}_{0}(=0)$ are mutually exclusive and each path $[M]$ contributes to the path integral by the transfer weight,
$\mathcal{W}[M]:=\prod_{T=0}^{N_0-1} {\sf W}^{(T+1\leftarrow T)}_{M_{j_{T+1}},M_{j_{T}}}.$

The so-called Malliavin weighting is the path functional which gives the relative, or log, sensitivity of this path weight to the infinitesimal external field:
\eqn{\label{eq:def-q}
 q[M]\equiv \left.\frac{\partial \log \mathcal{W}[M]}{\partial h_{\rm ext}}\right|_{h_{\rm ext}=0}.} 
Below we will show that the average linear sensitivity of any path-functional $\mathcal{A}[M]$ reads 
\beq
\left.\frac{\partial }{\partial h_{\rm ext}}E[\mathcal{A}[M]]
\right|_{h_{\rm ext}=0} 
= E[q[M] \mathcal{A}[M]]|_{h_{\rm ext}=0}.
\eeq
In fact using the formal linear expansion;
$$\mathcal{W}[M]=\mathcal{W}[M]_{h_{\rm ext}=0}
(1+q[M]h_{\rm ext}+\mathcal{O}({h_{\rm ext}}^2)),$$
we find
\beqa
&&\frac{\partial E[\mathcal{A}[M]]}{\partial h_{\rm ext}}
\cr &&=
\lim_{h_{\rm ext}\to 0}
\sum^{[M]}\mathcal{A}[M]\frac{\mathcal{W}[M]-\mathcal{W}[M]_{h_{\rm ext}=0}}{h_{\rm ext}}P_0^{(0)}(M_0)
\cr &&=
\sum^{[M]}\mathcal{A}[M] q[M] \mathcal{W}[M]_{h_{\rm ext}=0}P_0^{(0)}(M_0),
\eeqa
where the last line on the r.h.s. is the expectation of $\mathcal{A}[M]q[M].$

To calculate $q[M]$ we recall the form $\mathcal{W}[M]:=\prod_{T=0}^{N_0-1} {\sf W}^{(T+1\leftarrow T)}_{M_{j_{T+1}},M_{j_{T}}}.$ Using the additivity of the log of product,
we have
\beq
q[M]=\sum_{0\le T\le N_0-1}^{[M]}\left.\frac{\partial \log W_{M_{T+1},M_{T}}^{(T+1\leftarrow T)}}{\partial h_{\rm ext}}\right|_{h_{\rm ext}=0},
\eeq
where the sum is taken along the history $[M]$.
Therefore, the weight $q[M]$ can be calculated cumulatively along the process $M$.
Especially when the  perturbation is given uniquely at the stage-$(T_0-1),$ as in the main text, the relative sensitivity is reduced to $q[M]=\partial \log [W^{(T_0\leftarrow T_0-1)}_{M_{T_0},M_{T_0-1}}]/\partial h_{\rm ext}|_{h_{\rm ext}=0}.$
In fact if we regard the r.h.s. of Eq.(\ref{eq:dPdh}) as a path integral, 
the contribution of the path $[M]$ reads $\mathcal{W}[M] q[M].$

\section{Construction of final distribution from early stage one using martingale conditional expectation \label{sec:appendixC}}
For the simplicity of notations, we introduce (see (\ref{eq:geomopt}))
$$\mu_i={-T_0+2i}$$ 
$$ m_i=m^{\rm (eq)}_{T_0,\mu_i}$$
$$x_i=\mu_i+(N_0-T_0) \,m_i, \qquad i=0,1,\ldots,T_0$$
We will make up the final probability density $p(x)$ so that its normalization is
$\int_{x_0}^{x_{T_0}}p(x)dx=1.$ We suppose that $p(x)$ is piecewise linear whose
joint-points are $\{x_i,p(x_i)\}.$ The normalization condition then reads
{\small
\beqa 1&=&\!\!\!\sum_{i=0}^{T_0-1}\frac{p(x_i)+p(x_{i+1})}{2}(x_{i+1}-x_i)
\cr &=& p(x_0)\frac{x_{1}-x_0}{2}
+\!\!\!\sum_{i=1}^{T_0-1}p(x_i)\frac{x_{i+1}-x_{i-1}}{2}
+p(x_{T_0})\frac{x_{T_0}-x_{T_0-1}}{2}. 
\cr &&
\eeqa}
Then we define $p(x_i)$ through
\beqa
p(x_0)\frac{x_{1}-x_0}{2}&=&P^{(T_0)}_{h_{\rm ext}}(m_0),
\cr p(x_i)\frac{x_{i+1}-x_{i-1}}{2}&=& P^{(T_0)}_{h_{\rm ext}}(m_i) \qquad i=1,\ldots,T_0-1
\cr p(x_{T_0})\frac{x_{T_0}-x_{T_0-1}}{2}&=&P^{(T_0)}_{h_{\rm ext}}(m_{T_0})
\eeqa 
so that the ``ray'' of geometrical optics carries the probability from $T=T_0$ to 
$T=N_0.$ The martingale prediction of the probability densities in Fig.\ref{fig:fig3} are thus made.

\bibliographystyle{apsrev4-2.bst}
    \bibliography{ken_LNP_sar}
\end{document}